# Astronomical Symbolism in Australian Aboriginal Rock Art

Ray P. Norris[1] and Duane W. Hamacher[1]

[1]Department of Indigenous Studies, Macquarie University, NSW 2109, Australia
Corresponding author e-mail: ray.norris@csiro.au

**Abstract**

Traditional Aboriginal Australian cultures include a significant astronomical component, perpetuated through oral tradition and ceremony. This knowledge has practical navigational and calendrical functions, and sometimes extends to a deep understanding of the motion of objects in the sky. Here we explore whether this astronomical tradition is reflected in the rock art of Aboriginal Australians. We find several plausible examples of depictions of astronomical figures and symbols, and also evidence that astronomical observations were used to set out stone arrangements. However, we recognise that the case is not yet strong enough to make an unequivocal statement, and describe our plans for further research.

1.  **Introduction**

The dark night skies of Australia are an important part of the landscape, and would have been very obvious to Aboriginal people around their campfires before the British occupation. So it is unsurprising to find that stories of the Sun, Moon, planets, and constellations occupy a significant place in the oral traditions of Aboriginal Australians. This was first described by Stanbridge (1857), and since noted by many other authors (e.g. Mountford 1976; Haynes 1992; Johnson 1998; Cairns & Harney 2003; and Norris & Norris 2009). The focus of most of these works is on the correspondence between constellations, or celestial bodies, and events or characters in traditional Aboriginal oral traditions.

For example, in many Aboriginal cultures the European constellation of Orion is associated with young men, particularly those who are hunting or fishing (e.g. Massola 1968; Wells 1973). The constellation is called Djulpan in the Yolngu language, and the three stars of Orion's belt are associated with three brothers sitting across the width of a canoe, with Betelgeuse marking the front of the canoe, and Rigel the back. The three brothers were blown into the sky after one of them had illegally captured a kingfish, which corresponds to Orion's sword (*ibid*).

Similarly, the cluster of stars known to Europeans as the Pleiades, or Seven Sisters, are associated in many Aboriginal cultures with a group of young girls, or sisters (e.g. Massola 1968; Harney 1959; Andrews 2005). Many traditional Aboriginal stories refer to the sisters as pursued by the young men in Orion, which is curiously similar to the traditional European myth about these constellations. This may indicate either cultural convergent evolution, reflecting the subjective masculine and feminine appearance of Orion and the Pleiades respectively, or else suggest a much earlier story common to both cultural roots.

In addition to these narratives, the skies had practical applications for navigation and time keeping (e.g. Cairns & Harney 2003; Clark 1997). It has also been argued by Norris & Hamacher (2009) that a deep intellectual content is also present, in which meaning is sought for astronomical phenomena such as eclipses, planetary motions, and tides.

An impediment to this study is the widespread but mistaken belief that "No Australian Aboriginal language has a word for a number higher than four" (Blake 1981) which persists even though complex Aboriginal number systems have been well-documented in the literature (e.g. Tindale 1925; Harris 1987; McRoberts 1990; Tully 1997). Such colonial belief systems also maintain the misconception that Aboriginal people would not be interested in or capable of careful astronomical measurements, but we find no evidence to support this belief.

In this paper, we explore the extent to which these strong astronomical traditions are reflected in Aboriginal rock art. It is instructive to note that in some cases where the traditional Aboriginal culture is largely intact, we have first–hand accounts linking rock-art to astronomy (e.g. Cairns & Harney 2003). However, in some of these cases, such as Figure 1, the astronomical connection may not be apparent to a Western researcher unless guided by cultural knowledge.

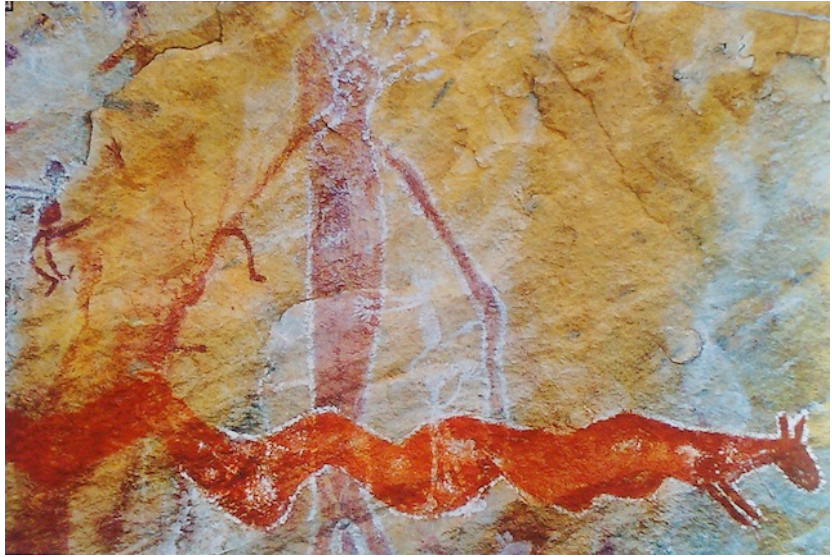

**Figure 1:** Wardaman rock painting of the "Sky Boss" and the Rainbow Serpent. The serpent at the bottom represents the Milky Way, and the head of the Sky Boss is associated with the "Coalsack nebula", although a researcher could not deduce this astronomical connection without access to the cultural insight of Wardaman elder Bill Yidumduma Harney. (Photo courtesy of Bill Yidumduma Harney)

Other rock art sites in Australia display motifs of astronomical designs. For example, on a hilltop near Palm Valley, Northern Territory is a depiction of the sun, crescent moon, and stars (Austin-Broos 1994). However, in this case the Arrernte description of the site has evolved from a traditional view to one that was shaped by Lutheran missionaries that settled the area in the late 19$^{th}$ century. When asked about the purpose of these astronomical engravings, the Arrernte informant explained that they had been placed there by God to direct Jesus while in the desert. Many features of the landscape had incorporated Christian mythology into the pre-existing oral traditions, including that of a star that was said to have fallen and made a hole between two trees where the Hermannsburg church was later built. Other geographical features of the area served a mnemonic purpose relating the land to Christian events, such as Noah's Flood and the inception of the 10 commandments (ibid).

Near Kalumburu, Western Australia is a rock painting on the side of a rock called Comet Rock. Bryant (2001) suggests that the rock painting represents a motif of a cometary fragment that impacted the Indian Ocean causing a great tsunami that swept over the land, which he speculates is supported by Aboriginal stories. The rock is 5 km from the ocean on a plain covered in a layer of beach sand. A similar example is given by Jones (1989) who describes a cosmic impact, followed by a deluge, in the Darling Riverbed near Wilcannia, NSW. According to Jones, rock art on Mt Grenfell showing people standing on one another's shoulders may represent people climbing the mountain to escape the flood.

In addition, certain star-like motifs can be found engraved in rock in the Sydney-Hawkesbury region (e.g. Sim 1966), although their meaning is unknown. These designs typically involve a small circle with lines radiating outward, suggesting a sun, star, or "sunburst" motif.

2.    **The Sun and Moon in Aboriginal Rock Art**

In most Aboriginal cultures, the Moon is male and the Sun is female. For example, a Yolngu oral tradition explains the motion of the Sun in terms of Walu, the Sun-woman. She lights a small fire each morning, producing the dawn (Wells 1964), and decorates herself with red ochre, some of which spills onto the clouds, to create the red sunrise. Carrying a blazing torch made from a stringy-bark tree, she travels across the sky from east to west, creating daylight. At the western horizon, she extinguishes her torch, and travels back underground to her morning camp in the east. Warner (1937) reports that he was told "the Sun goes clear around the world" by a Yolngu man who illustrated this "by putting his hand over a box and under it and around again".

The Yolngu people call the Moon-man Ngalindi. The phases of the Moon are caused by Ngalindi being attacked by his wives, who chopped bits off him with their axes, reducing him from the fat full moon to the thin waning Moon (Wells 1964; Hulley 1996), and eventually dying (the new Moon). After staying dead for three days, he rose again, once more growing round and fat to become the full Moon, when his wives attacked him again.

Yolngu culture also recognises that the tides are caused by the Moon, and that the height of the tides depends on the phases of the Moon. This is explained in terms of a complicated interaction between the rising Moon and the Sea, the Moon alternately filling and emptying, depending on its phase, as it rises through the ocean horizon. While it is dangerous to generalise from one Aboriginal culture (Yolngu) to others, there exist similarities that transcend Aboriginal cultures, such as the gender of the Sun and Moon, which are almost universally female and male respectively.

Given these strong oral traditions, we might expect to find depictions of the Sun and Moon in Aboriginal rock art. Obvious examples of solar images exist, such as those at Ngaut Ngaut, SA (Figure 2) and many more are surmised, such as the "bicycle wheel" or "sunburst" petroglyphs in the Panaramittee engravings at Sturts Meadows, NSW (Figure 3). However, the latter can entertain many interpretations, including a supernova (Murdin 1981), and caution is required when interpreting such images in the absence of cultural context.

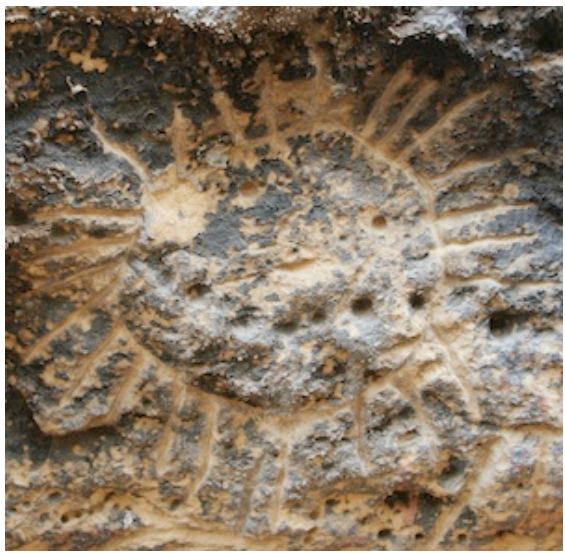

**Figure 2:** The Sun engraving at Ngaut Ngaut, South Australia.

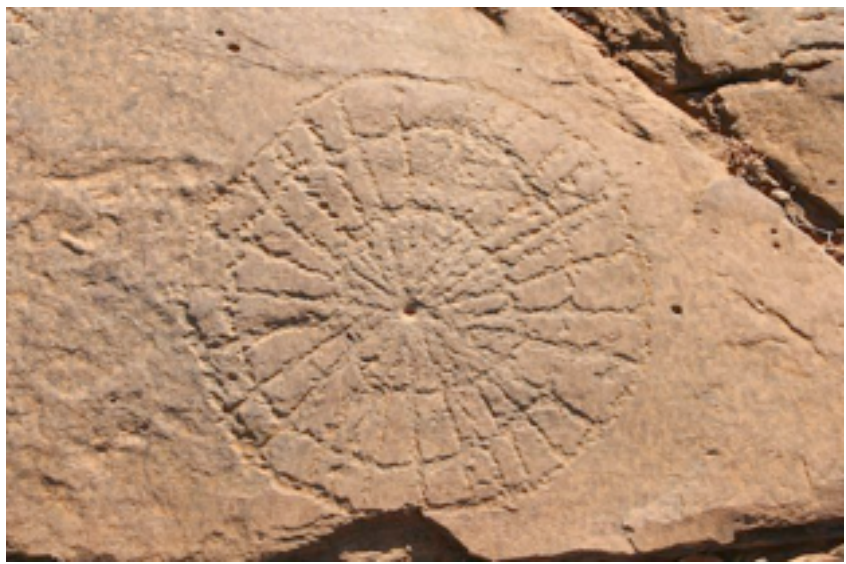

**Figure 3:** A "bicycle-wheel" or "sunburst" petroglyph at Sturts Meadows, NSW. While this may represent the sun, or perhaps even a supernova, there is no additional information to support these interpretations, and so any interpretation remains speculative.

Crescent shapes are also common, and may represent the moon, although they have also been attributed to boomerangs. Many examples of crescent shapes are found in the Sydney Basin rock engravings, and are traditionally referred to as boomerangs (e.g. McCarthy 1983). However, there is a clear difference between a boomerang-shape and a crescent moon: boomerangs typically have straight sides and rounded ends, whereas the crescent moon always has a curved shape and pointed ends (Figure 4). It is therefore likely that the Sydney Rock Engravings contain a significant astronomical component.

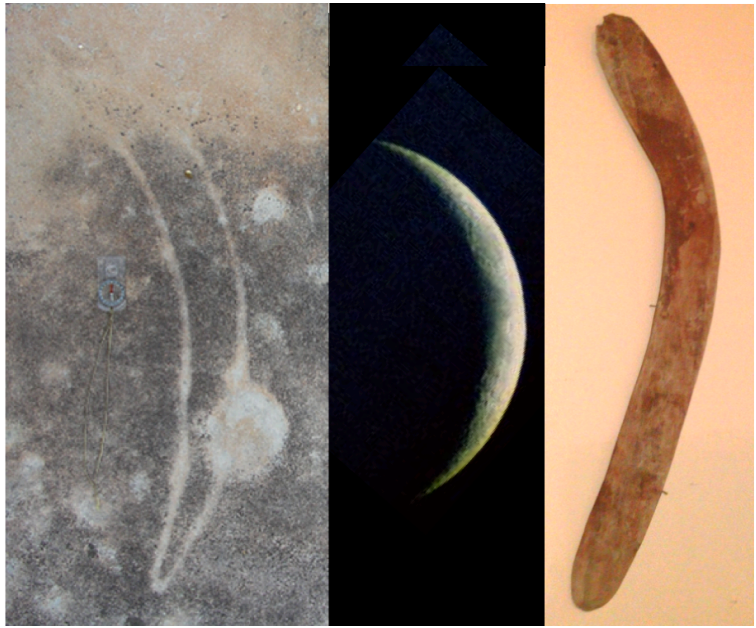

**Figure 4:** A crescent-shaped rock engraving from Calga Springs, NSW; the crescent Moon, and a Boomerang. Note the smooth curve and pointed ends of the engraving, as opposed to the relatively straight lines and rounded ends of the boomerang.

3. **Eclipses**

Several Aboriginal cultures recognised that eclipses are caused by a conjunction of the Sun and Moon. For example, in Arnhem Land a solar eclipse is caused by the Sun-woman being hidden by the Moon-man as they make love, while a lunar eclipse is caused when the Moon-man is pursued and caught by the Sun-woman (Johnson 1998; Warner 1937). Similarly, Bates (1944) reports that the solar eclipse of 1922 was said by the Wirangu people to be caused by the Sun and Moon "becoming husband and wife together".

These stories demonstrate a significant intellectual accomplishment. To understand a solar eclipse, in which the Moon comes between the Earth and the Sun, is impressive, but perhaps not surprising if traditional Aboriginal thinkers carefully studied the motion of the Sun and Moon. However, to understand a lunar eclipse, in which the Earth's shadow extinguishes the Moon, requires a significant leap of understanding, since it occurs when the Sun and Moon are diametrically opposed in the sky, and it is precisely this alignment that causes the eclipse.

Amongst the crescent motifs found in the Sydney Rock Engravings are several examples that depict a man and woman under, or next to, a crescent shape. In the case of the Basin Track engraving (Figure 5), the sign erected by the National Parks and Wildlife Service explains that the engraving shows a man and woman with a boomerang. However, it is unclear why a man and woman should reach up towards a boomerang in the sky. The engraving makes more sense if, as suggested above, the crescent represents a Moon rather than a boomerang, but is still unusual in that the moon is shown with the two horns pointing down, a configuration normally seen only when the moon is barely visible in the early morning or late afternoon. But such a configuration can be seen during an eclipse, and this suggestion is supported by the two figures, one of which partially obscures the other. Such carefully drawn obscuration is unusual in these rock carvings, and in this case may represent the Moon-man obscuring the Sun-woman (or vice-versa) during an eclipse. Other circumstantial evidence for this hypothesis is that the man and woman face toward the northeastern horizon, in the direction a solar eclipse could be seen in the early morning. For example, such an eclipse, with the horns pointing downwards as depicted, took place on the morning of 8 August 1831, although it is not clear whether any Guringai people still lived in the area at that time. John Clegg (personal communication) has also speculated that a hermaphrodite figure near this engraving may represent the moon-man and sun-woman fully superimposed during a total eclipse.

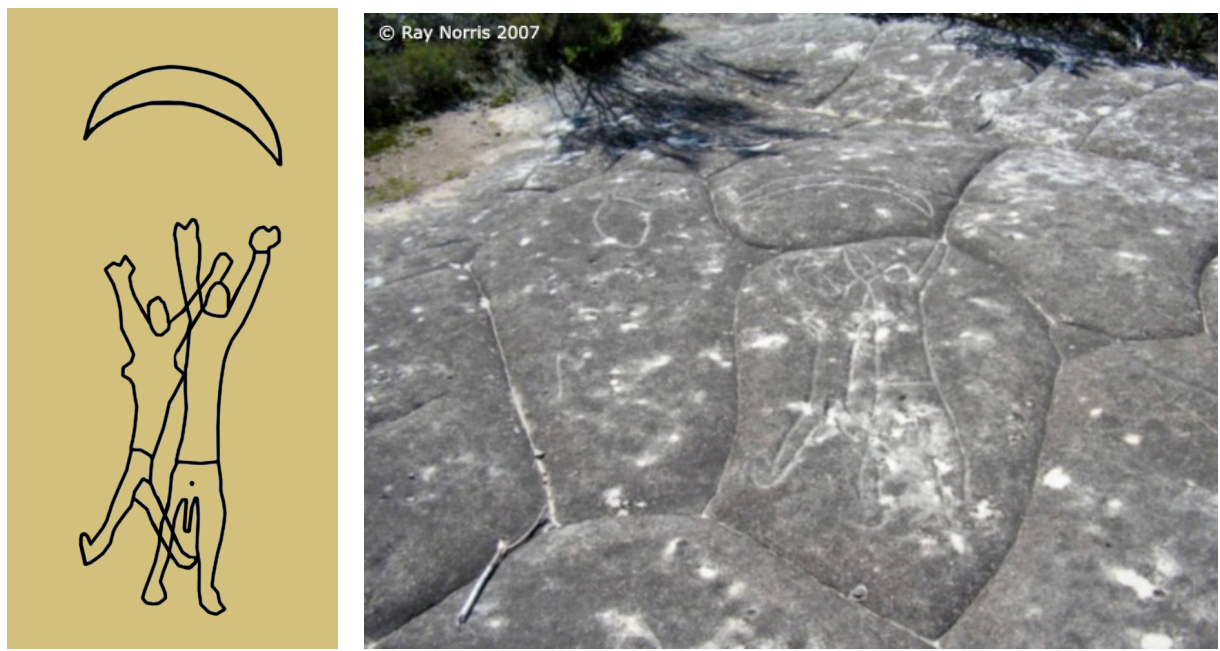

**Figure 5:** The "eclipse" engraving at the Basin Track, Ku-ring-gai Chase National Park, NSW.

4.      **The Emu in the Sky**

Across the zenith of an Australian Autumn sky stretches the bright band of the Milky Way. Within it can be seen a number of dark patches and streaks, caused by clouds of interstellar dust in which new stars are being born. Perhaps the best-known Aboriginal constellation is the "Emu in the Sky", formed not of stars, but of the dark patches between them. Amongst the Sydney Rock Engravings, close to the Elvina Track, is a finely engraved emu (Figure 6), carefully drawn to show the gizzard and other anatomical features. On the other hand, its legs trail behind it, in a position that would be unnatural for a real emu, but is very similar to that of the Emu in the Sky. This was first noted by Cairns (1996), who suggested the engraving might represent the Emu in the Sky rather than a real emu. This suggestion is further supported by the fact that the time of the year when the Emu in the Sky stands in the evening above her portrait, in the correct orientation, is the same time when real-life emus are laying their eggs. It seems quite possible that this engraving is a picture of the Emu in the Sky rather than a real emu.

5.      **Astronomical Records**

Figure 2 shows one of the engravings of the Sun and Moon at Ngaut Ngaut, SA, providing clear evidence of an astronomical connection at this site. Close to this engraving are carved a series of dots and lines (Figure 7), which, according to the traditional Nganguraku owners, show the "cycles of the Moon". This oral tradition has been passed through generations from father to son, but since initiation ceremonies were banned (along with the Nganguraku language) by Christian missionaries over a hundred years ago, only this fragment of knowledge survives, and it is not known exactly what the symbols mean. Certainly the dots and lines resemble tally marks, and the site's astronomical connection suggests that they may indeed represent astronomical records. However, we have so far failed to decode them. We plan to conduct further tests to search for evidence of astronomical periodicities in the marks.

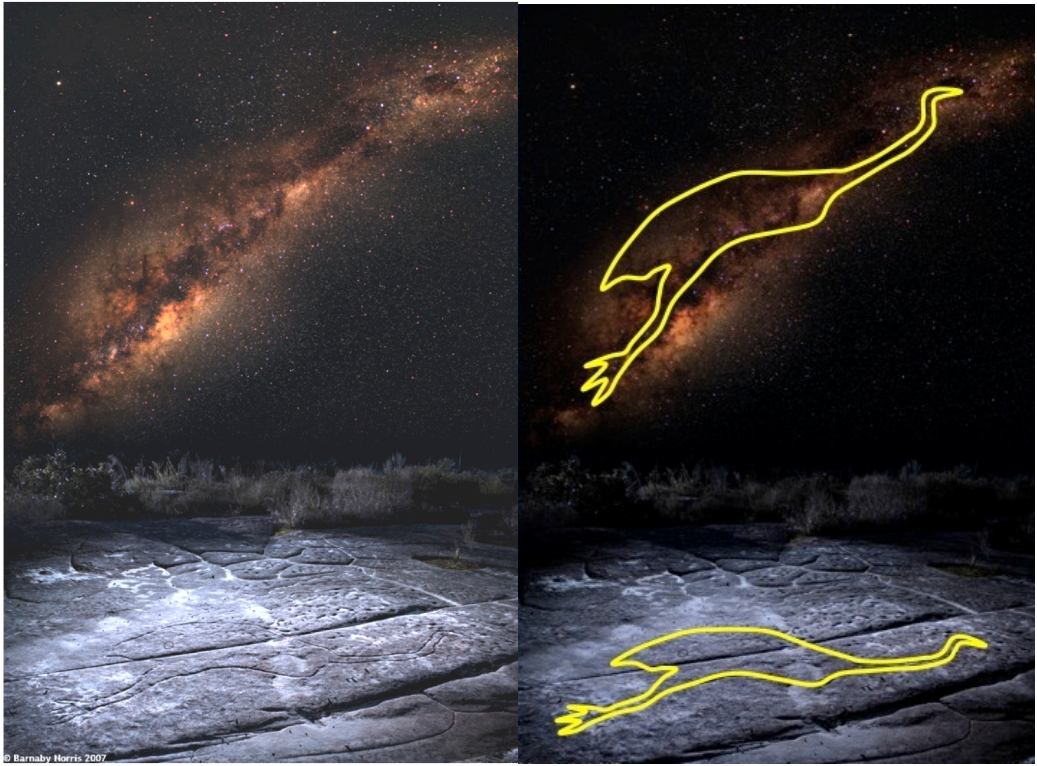

**Figure 6:** The "Emu in the Sky", consisting of dark patches in the Milky Way, above the Emu engraving at Elvina track, Ku-ring-gai Chase National Park, NSW. Photo by Barnaby Norris (2007).

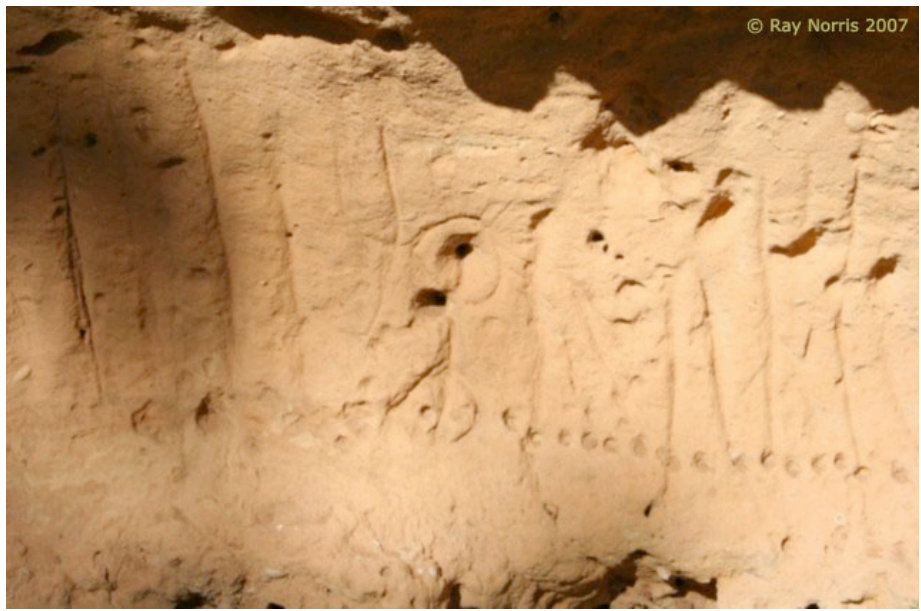

**Figure 7:** The Ngaut Ngaut rock art site near the banks of the Murray River north of Adelaide, South Australia.

### 6. Stone Arrangements and Bora Rings

Many stone arrangements and Bora (ceremonial) rings, throughout Australia, appear to be aligned to the cardinal points. For example, the Carisbrook site in Victoria (Figure 8) is a boomerang-shaped (as opposed to a lunar crescent) arrangement whose axes point within a few degrees of due north and due east.

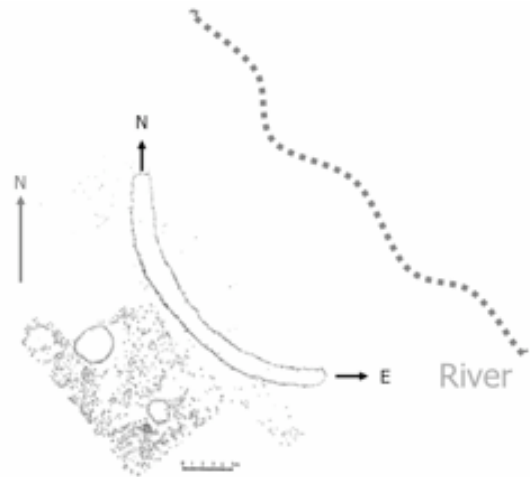

**Figure 8:** Plan of the Carisbrook stone arrangement, Vic. Adapted from Lane & Fullagar, 1980

Since there is no "pole star" in the Southern Sky, finding the cardinal points without a compass or GPS is not easy. The well-known technique of plotting lines from the Southern Cross and pointers is only useful if someone else has already determined true South and then tells you how to find it, and so leaves unresolved the question of how to determine the position of true South in the first place. One possible technique is to observe the rotation of the Southern Cross over the course of a winter's night, mark the position on the horizon vertically below its extreme easterly and westerly positions, then mark the half-way point between them. Similarly, east and west may be found by marking the extremes of the rising or setting sun's locations on the horizon over the year, then marking the halfway point between them. Either process requires a process of astronomical observation and measurement, and the intellectual motivation to do so. Thus the very existence of a significant number of structures aligned to the cardinal points implies a degree of planning, observation, and measurement that seem to be absent from most anthropological accounts of Aboriginal cultures. Of course, they could be aligned to the cardinal points purely by chance, which implies that there must be a far larger number of structures that are <u>not</u> aligned to cardinal points. While experience suggests that this is not the case, a proper statistical study is currently in planning stage by Hamacher & Norris.

A notable example, the Wurdi Youang stone arrangement built by the Wathaurung people in Victoria, is an egg-shaped ring of stones, about 50m in diameter, with a major axis almost exactly east-west (Figure 9). At its Western apex are three prominent waist-high stones mimicking three mountains in the distance. Morieson (2003) has suggested that some outlying stones to the West of the circle indicate the setting positions of the Sun at the equinoxes and solstices, when viewed from the three large stones. Norris et al (2010) have confirmed these alignments. More importantly, Norris et al. have shown that the straight sides of the circle also indicate the solstices, and are parallel to the lines to the outliers proposed by Morieson. Thus at this site we have two independent sets of indicators, both indicating the same positions on the horizon, corresponding to the Sunset position at the two solstices and the equinox. Taken at face value, this suggests that astronomical observations were used to construct the site, and may have been important in its use.

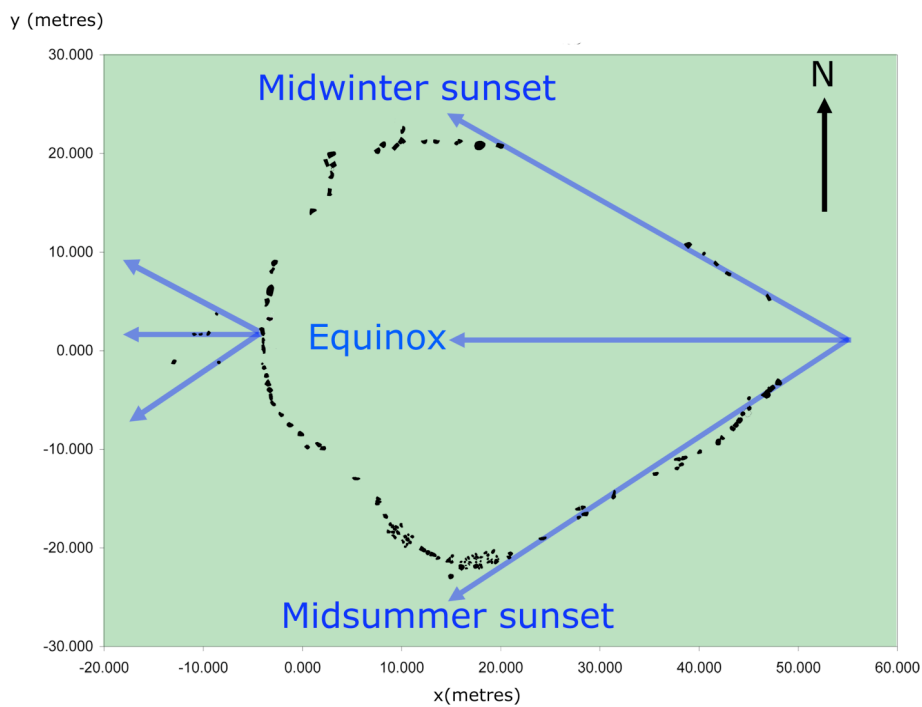

**Figure 9:** Plan of Wurdi Youang stone arrangement, VIC. (From Norris et al 2010).

However, the apparent significance of this site is still subject to some doubts:

1) The outliers are only accurate to a few degrees - could these alignments have occurred by chance?

2) Although the stones of the circle are large and immovable, the outliers are small and could have been moved.

3) There are other stones in the vicinity, requiring a subjective decision as to which stones to include in a survey.

These doubts are partly addressed by the alignments of the straight lines in the egg-shape, but an even better way to address them would be to find other sites with similar astronomical alignments. This is a subject of continuing research.

## 7. Conclusion

It is well established that traditional Aboriginal oral traditions include significant references to the sky, and the motion of the bodies within it. It is becoming clear that this knowledge includes a deep understanding of the motion of celestial bodies, and so it should not come as a surprise to find this knowledge reflected in rock art.

In this paper we provide plausible examples in which astronomical figures may be depicted in rock art, but we are certainly not yet in the position where we can claim to have provided conclusive evidence. We hope that such a case may be made in the near future, based in part on statistical data. However, a sufficient body of circumstantial evidence is now accumulating It is therefore important to take such astronomical oral traditions into account when trying to understand the significance of rock art.

## 8. Acknowledgements

This project is dedicated to the hundreds of thousands of Indigenous Australians who lost their lives after the British occupation of Australia in 1788. We are grateful to the Indigenous groups who have welcomed us onto their land and given us permission to tell their stories. We also thank Barnaby Norris and Cilla Norris, who helped with much of the research described here, and our colleagues Hugh Cairns, John Clegg, Paul Curnow, Kristina Everett, and John Morieson for their help and collaboration.